# Carbon-doped high mobility two-dimensional hole gases on (110) faced GaAs

S. Schmult, C. Gerl, U. Wurstbauer, C. Mitzkus and W. Wegscheider

Email: stefan.schmult@physik.uni-regensburg.de

Universität Regensburg, Institut für Experimentelle & Angewandte Physik, D-93040 Regensburg, Germany

Carbon-doped high mobility two-dimensional hole gases grown on (110) oriented GaAs substrates have been grown with hole mobilities exceeding $10^6 cm^2/Vs$ in single heterojunction GaAs/AlGaAs structures. At these high mobilities, a pronounced mobility anisotropy has been observed. Rashba induced spin-splitting in these asymmetric structures has been found to be independent on the transport direction.

Recently, carbon-doped high mobility two-dimensional hole gases (2DHGs) in GaAs/AlGaAs heterostructures grown along the [001] direction have been presented, employing carbon as acceptor [1-3]. Compared to previous structures obtained using beryllium as an acceptor, carbon has been proven to solve problems associated with impurity segregation and diffusion. For applications in Spintronics the use of the (110) surface was suggested. Here, spin relaxation processes are partially absent, and thus, spin lifetimes can become large [4]. Employing silicon as an acceptor, 2DHGs have been presented on (110) GaAs recently [5]. The achievement of these systems allows in the future the realization of structures based on the cleaved edge overgrowth [6] and a combination of this technique and spintronic devices. Unfortunately, high-mobility growth on nonpolar (110) oriented GaAs requires very stringent growth conditions which are usually accompanied with deteriorated transport properties compared to their (001) oriented counterparts.

In this letter, we report on results of GaAs bulk p-doping on the (110) GaAs surface employing carbon, as well as the realization of high mobility two-dimensional hole gases (2DHG) with hole mobilities exceeding $10^6 cm^2/Vs$ at a density of $1.25*10^{11} cm^{-2}$ on this assigned surface. In a L-shaped Hallbar geometry, a record hole mobility of $1.13*10^6 cm^2/Vs$ along the [-110] direction has been observed, in contrast to $4.8*10^5 cm^2/Vs$ along the [001] direction. In addition, 2DHGs have been grown on cleaved edges of GaAs [110] and [001] wafers, in order to determine the transport properties along the [001] and [-110] direction. The results are in quantitative agreement with our findings on (110) substrates. Structure inversion asymmetry induced spin-splitting of the heavy hole subband has been observed at low temperatures. An analysis does not reveal a difference in the spin-split densities for the two investigated transport directions.

A sketch of the modulation doped single interface (MDSI) structure is given in Fig. 1. The samples have been grown in a UHV MBE system, achieving electron mobilities over $10^7 cm^2/Vs$, utilizing a novel carbon filament source [7]. On a semi-insulating (110) GaAs substrate, first a superlattice is grown, followed by a 500nm thick GaAs layer, an $Al_{0.3}Ga_{0.7}As$ spacer layer of 60nm thickness, a carbon-δ-doping layer, a 90nm thick $Al_{0.3}Ga_{0.7}As$ layer and a 10nm GaAs cap layer. The growth conditions correspond to these for (110) high mobility 2DEGs [8]. Bulk doping with a filament power of 470 W (including parasitic dissipation due to serial resistances) resulted in a hole density of $3*10^{18} cm^{-3}$, determined to be 30% as efficient as on the (001) surface. On the 2DHGs grown on (110) substrates, L-shaped Hallbars have been prepared. The two arms of the L-shaped Hallbar, oriented in [001] and [-110] directions, are 2mm long and 400µm wide. For the growth on cleaved edges, 500µm thick [001] and [110] substrates have been cleaved across the non-polar (110) and (1-10) surfaces respectively. The prepared surfaces show no macroscopic steps if inspected with Nomarski microscopy. Immediately after cleaving, the prepared substrates have been transferred into the MBE system. After growth using the identical layer sequence as described above, four contacts are established on the cleaved edges, allowing measurements of the longitudinal voltage in four-terminal geometry (Fig.1). Measurements at temperatures of 1.3K and 30mK have been carried out in a $^4$He system and in a $^3$He/$^4$He dilution refrigerator, respectively. The measurements are performed with standard lock-in technique, applying currents between 20nA and 1µA. Here, densities and mobilities are measured in the dark. We explicitly note this fact, since illumination dependent conductivity was observed in all samples and density tuning is possible using light. In contrast to high mobility hole gases on (001) GaAs, persistent photoconductivity has been observed, critically depending on the temperature during illumination.

For a 2DHG on a (110) substrate, equipped with an L-shaped Hallbar, Hall and longitudinal voltages for a temperature of 30mK are plotted in Fig. 2 (A) along the [-110] transport direction. For a density of p=1.25*10$^{11}$cm$^{-2}$, hole mobilities of 1.13*10$^6$cm$^2$/Vs (|| [-110]) and 4.8*10$^5$cm$^2$/Vs (|| [001]) are

determined. In Fig. 3, the low magnetic field behaviour, exhibiting pronounced beatings, is displayed for both directions (a and b). From these data, two densities of $p_1=0.56*10^{11}cm^{-2}$ and $p_2=0.69*10^{11}cm^{-2}$ can be extracted for both directions, attributed to the densities of the two spin-split heavy hole subbands. The huge difference in density of 23% can be explained by the strongly non-symmetric structure resulting from one-sided doping.

For the 2DHGs grown on cleaved edges of (001) (transport along the [-110] direction) and (110) substrates (transport along the [001] direction), mobility values at a temperature of 1.3K of $5.6*10^5 cm^2/Vs$ at densities $1.25*10^{11}cm^{-2}$ and $3.7*10^5 cm^2/Vs$ at $1.7*10^{11}cm^{-2}$ respectively have been determined. These results are consistent with the findings for the L-shaped Hallbar measurements mentioned above. The lower mobilities compared to the values for the 2DHG on the (110) substrate are attributed to non-optimal growth conditions for cleaved edge growth. Additionally, for the high mobility cleaved edge sample, measurements of the longitudinal resistance at 30mK have been carried out in order to determine the low temperature mobility and to reveal the heavy hole spin-splitting. The result is given in Fig. 2 (B). The mobility increases up to $6.6*10^5 cm^2/Vs$, in contrast to $5.6*10^5 cm^2/Vs$ at 1.3K. Fig. 3 (c) shows the low magnetic field Shubnikov-de-Haas oscillations, which reveal after periodicity analysis the two heavy hole subband densities of $p_1=0.56*10^{11} cm^{-2}$ and $p_2=0.69*10^{11} cm^{-2}$. These are the same values as reported above for the 2DHG grown on a (110) substrate.

Compared to two-dimensional electron systems, for hole systems roughly half of the density can be achieved for similar spacer layers. The reason is the lower value of the valence band discontinuity, in comparison to the conduction band offset. Due to crossing of the $\Gamma$ and X bands, an increase of the aluminium mole fraction above 33% results in higher electron densities, but not in higher electron mobilities. As a result of the different valence band structure this restriction does not apply for modulation doped hole systems. With an identical sample, but 50% aluminium mole fraction in the barriers, we tried to verify this for hole systems. At 1.2K, a van-der-Pauw mobility of $9.3*10^5 cm^2/Vs$ was measured at a density of $p=2.2*10^{11} cm^{-2}$. Another piece of this wafer was equipped with an aluminium top gate. Fig 4 shows the dependence of the hole mobility on the gate voltage tuned density, as well as the dependence of the density on the gate voltage at 1.2K. We note here, that the mobility follows perfectly $\mu \sim p^\alpha$, with $\alpha=1$, and saturates at a maximum mobility of $8*10^5 cm^2/Vs$.

In conclusion, carbon-doped high mobility two-dimensional hole gases in single heterojunction GaAs/AlGaAs structures have been fabricated on (110) faced GaAs substrates. The achieved hole mobility for the [-110] transport direction exceeds $10^6 cm^2/Vs$. This is remarkable as for (001) oriented 2DHS where much more favourable growth conditions can be employed, the mobility is lower for a comparable density [3]. A pronounced anisotropy of the mobility of more than 50% was observed. These results are consistent with mobility measurements in 2DHGs on cleaved edges of (001) and (110) GaAs. In the high mobility samples at low temperatures, spin-splitting results as a consequence of the structure inversion asymmetry, revealing equal quantities of spin-splitting along the investigated transport directions.

The authors like to thank Frank Fischer for fruitful discussions. The work was financially supported by the Deutsche Forschungsgemeinschaft (GK 638) and BMBF via the program "Spinelektronik und Spinoptoelektronik in Halbleitern".

Figure captions:

Fig. 1:

Sample structure of the investigated MDSI structures (upper portion) and schematic diagram of the ex-situ prepared (110) or equivalent growth template, using GaAs (001) (Arial) or *(110)* (*Italic*) substrates (lower portion). The relevant crystal directions are given. The wafers are cleaved outside the MBE system across the non-polar (110) or (1-10) surfaces respectively. After growth, the 2DHG on the top is equipped with four contacts. The current is applied between contacts 1 and 4, longitudinal voltage is measured across 2 and 3. The sketches are not to scale.

Fig. 2:

Hall and longitudinal voltages for the transport direction [-110] on (110) GaAs (A) and longitudinal voltage for the 2DHG on cleaved [001] substrate along the [-110] direction (B) at a temperature of 30mK.

Fig. 3:

Low magnetic field SdH oscillations at 30mK of 2DHG on (110) faced GaAs along the [-110] direction (a), along the [001] direction (b) and along the [-110] direction on cleaved [001] GaAs (c). In all cases pronounced beating pattern are clearly visible.

Fig. 4:

Mobility vs. density of a gated 2DHG with 50% aluminium mole fraction in the barriers and dependence of the density on the gate voltage at 1.2K. Up to a critical density the mobility rises linearly with density. Lines are a guide to the eye.

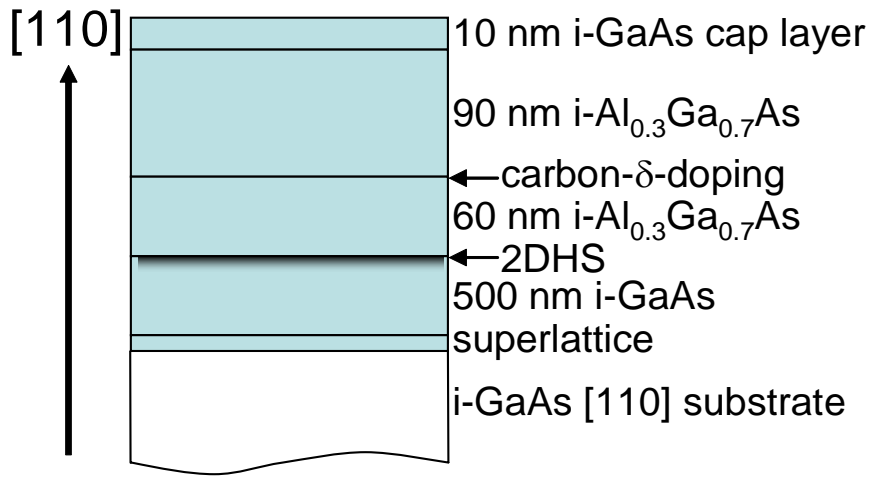
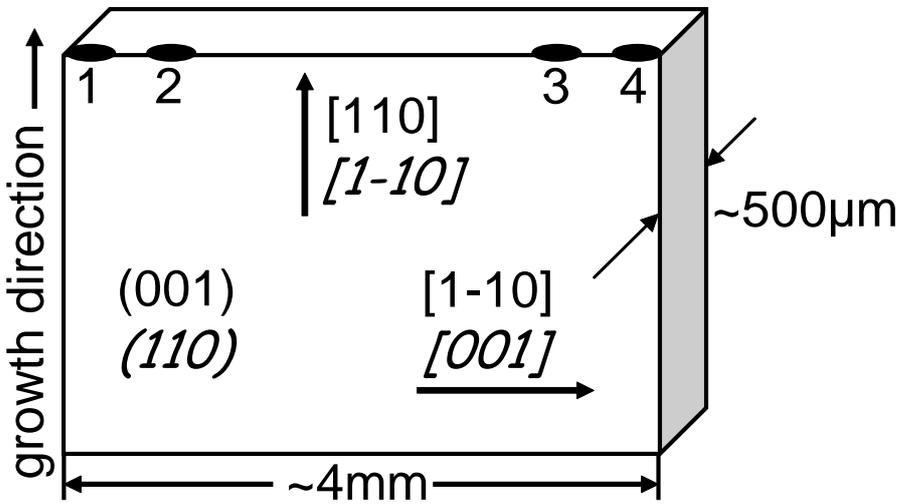

Fig. 1: S. Schmult et al.

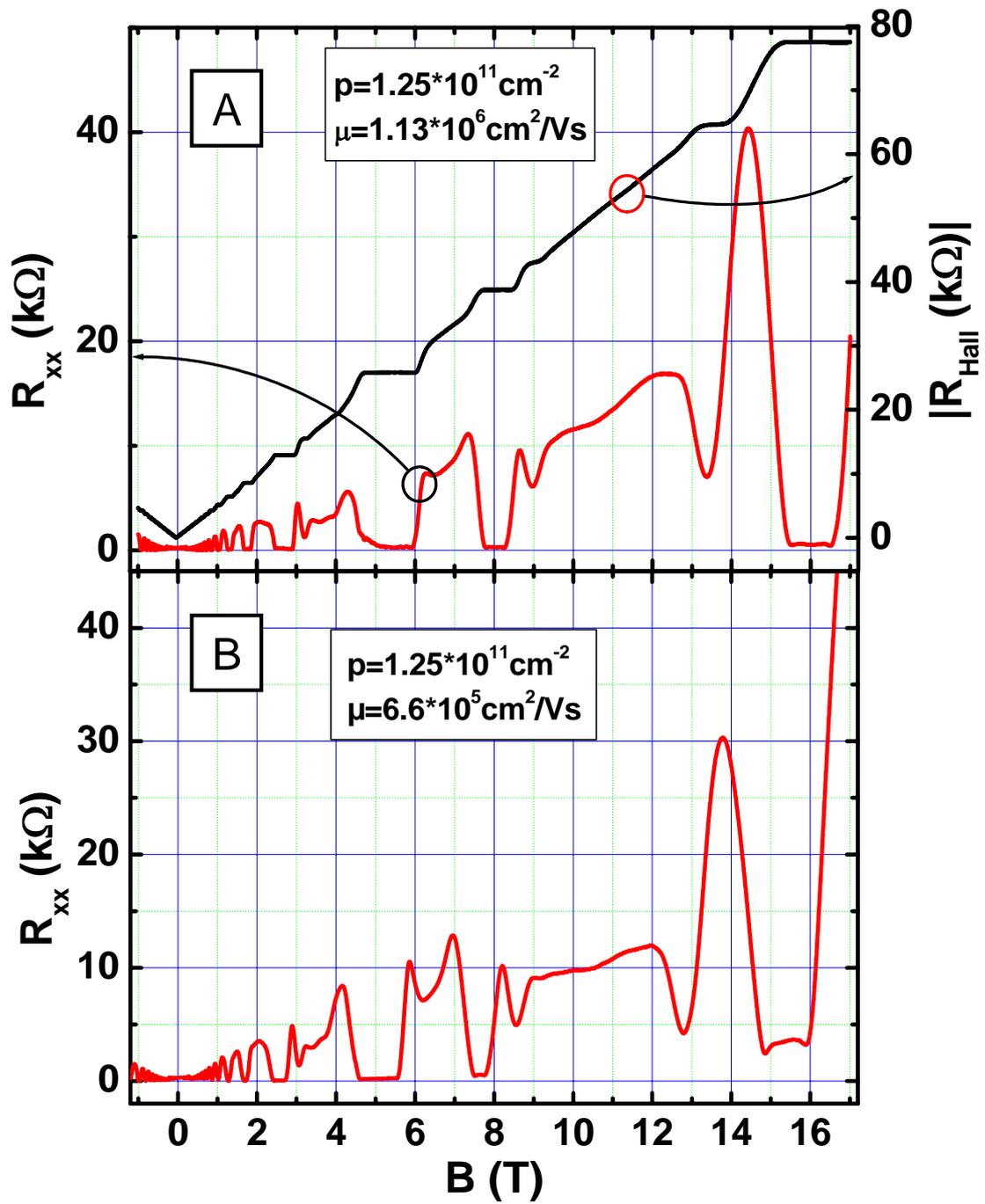

Fig. 2: S. Schmult et al.

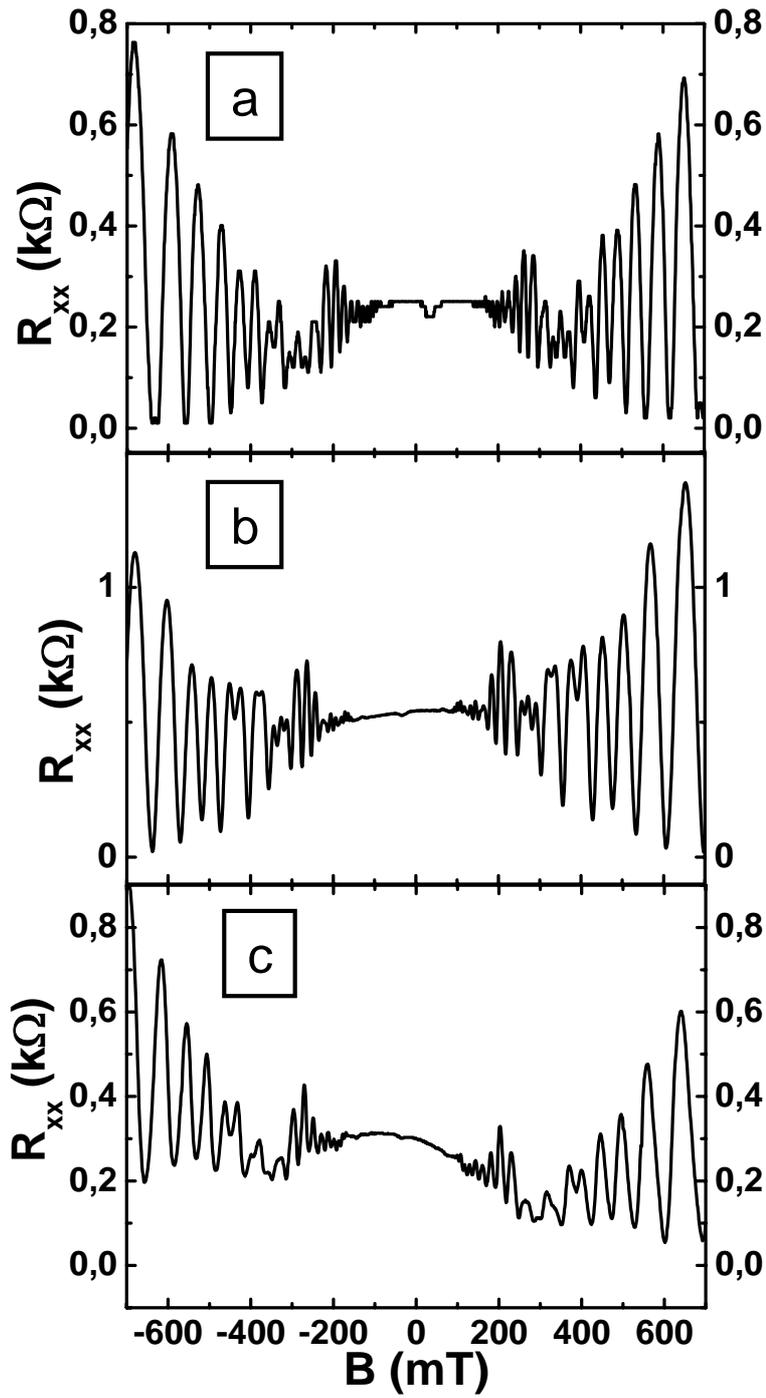

Fig. 3: S. Schmult et al.

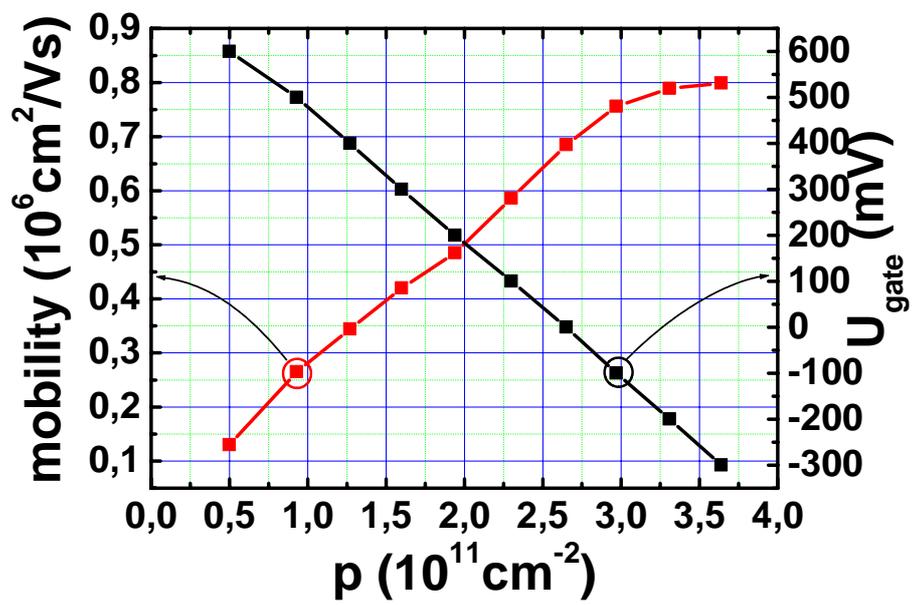

Fig. 4: S. Schmult et al.